# Assessment of a planetary model to predict Rieger periodicity in sunspots and flares

Ian R. Edmonds

**ABSTRACT**

This paper develops a planetary model to predict the occurrence of intermediate range periodicity in solar activity, in particular the ~155 day Rieger periodicity in flare activity. It is shown that periodicity at half integer multiples of the period of Mercury occurs consistently in indices of solar activity. For this reason the planetary model is based on the triggering of sunspot emergence when planetary tides peak at times of conjunction of Mercury with Venus, Earth and/or Jupiter. The periodicity of components in the planetary model match reasonably well the observed intermediate periodicity in sunspot and flare activity with the strongest model component occurring at 155 day period. A comparison of filtered versions of the model and the N.O.A.A. flare index at 155 day periodicity demonstrates the potential for both short term, (within a solar cycle), and longer term, (over several solar cycles), predictive capability of the model. However, this assessment finds that prediction of 155 day periodicity in flare activity is effective only when the north and south hemispheric components of flare activity are in-phase.

## 1 INTRODUCTION

The intermediate range of periodicity, 100 – 1000 days, in solar magnetic indices such as sunspot number, sunspot area and solar flares has been studied continuously since the observation, during solar cycle 21, of a strong ~155 day periodicity in solar flares by Rieger et al (1984). Subsequent studies have reported numerous periodicities in solar activity in the intermediate range (Dennis 1985; Lean & Bruekner 1989; Lean 1990; Oliver et al. 1998; Krivova & Solanki 2002; Tan & Cheng 2013; Choudhary et al. 2014; Chowdhury et al. 2015). Most attention has focused on periodicity in the 150 – 160 day range, the so-called Rieger periodicity, Ballester (2017). Discovering the origin of intermediate range periodicity is complicated by the fact that the observed periodicities are intermittent, being strong in some solar cycles and weak or non existent in other solar cycles. Occasionally, researchers have attempted to identify the origin of the periodicity. One idea is that there is some type of cycle or "clock" mechanism within the Sun itself and the observed periodicities are sub harmonics of the "clock" (Bai & Sturrock 1991; Bai & Sturrock 1993; Sturrock 1996; Krivova & Solanki 2002). Another idea is that the periodicities originate from Rossby type wave modes on the Sun, (Wolff 1992; Lou 2000; Zaqarashvili & Gurgenasvili 2018). An alternative idea is that the periodicities in solar activity are due, in significant part, to planetary influence on the Sun, (Bigg 1967; Wood 1972; Condon & Schmidt 1975; Kurochkin 1998; Hung 2007; Wolff & Patrone 2010; Wilson 2011; Abreu et al. 2012; Scaffetta 2012). However, the planetary approach has received very limited support, Charbonneau (2013), due, mainly, to the fact that planetary tidal effects appear to be too small to have any influence (de Jager & Versteegh 2005; Callebaut et al. 2012; Cameron & Schussler 2013).

In this paper we assess if a model based on planetary tidal influence can predict the observed periodicity in sunspot emergence and flares. Additionally, we test if the model predictions are consistent with the observed intermittency of intermediate range periodicity.

The paper is organised as follows. Section 2 outlines the data sources and analysis methods. Section 3 briefly describes the intermediate range periodicity observed in the sunspot area index, using the daily record extending from 1876 to 2012. Section 4 develops a quasi-physical model of the time dependence of sunspot emergence. Section 5 assesses the correspondence between the modelled and the observed intermediate range periodicity in sunspot emergence and compares the intermittency of the modelled periodicity with that of the observed periodicity. Section 6 compares observed and modelled intermediate periodicity in flares. Section 7 assesses whether the model has any predictive capacity in respect to the timing of sunspot emergence and the occurrence of flares. Section 8 assesses how the model predictions relate to solar activity on the north and south solar hemispheres. Section 9 is a discussion and conclusion.

## 2 DATA SOURCES AND ANALYSIS METHODS

The data used in this paper are records of daily sunspot area and flare activity.

Daily data for sunspot area north, south, and total from 1874 to present can was obtained from the NASA site: http://solarscience.msfc.nasa.gov/greenwch/daily_area.txt.
However, this paper uses the daily sunspot data for the 50,038 days from January 01, 1876 to December 31, 2012 to avoid gaps in earlier data. This interval includes solar cycles 12 to 23.

Data for north, south, and total flares was obtained from the comprehensive flare index at the NOAA site: https://www.ngdc.noaa.gov/stp/space-weather/solar-data/solar-features/solar-flares/index/comprehensive-flare-index/.
The data covers a 15,706 day interval January 01, 1966 to December 31, 2008, that includes solar cycles 20 to 23.

The data analysis methods used were the Fast Fourier Transforms (FFT) obtained using the FFT function available in the plotting application DPlot and simple running averages obtained using the SMOOTH function in DPlot. Where smoothed data appears in the figures the extent of the smoothing is indicated by the symbol Sxxx. For example data smoothed with a 365 day running average is denoted with S365. Band pass filtering of data by Inverse Fourier Transform was performed by first obtaining a FFT of the data then summing the cosinusoids associated with each spectral point occurring in some specified frequency band. Where a very narrow band filter is used or when the time variation is generated from only a few spectral components the frequency, amplitude and phase of each component is specified in the text. This is done in order to facilitate reproduction of the results by interested readers. Very broadband filtering is achieved by differencing two running averages of the daily data.

Planetary coordinates were obtained from the NASA COHOWEB site: http://omniweb.gsfc.nasa.gov/coho/helios/planet.html . This site provides daily values of orbital radius and heliospheric longitude for each planet from January 01, 1959 to December 31, 2019.

## 3 THE OBSERVED INTERMEDIATE PERIODICITY IN SUNSPOT AREA

Figure 1 shows the low frequency range in the spectra obtained by FFT of daily sunspot area between 1876 and 2012. The spectra have been smoothed by a five point running average. Spectral peaks which are common to both the sunspot area north (SSAN) and the SSAS spectra are marked with vertical reference lines labelled with the period of the peaks in days.

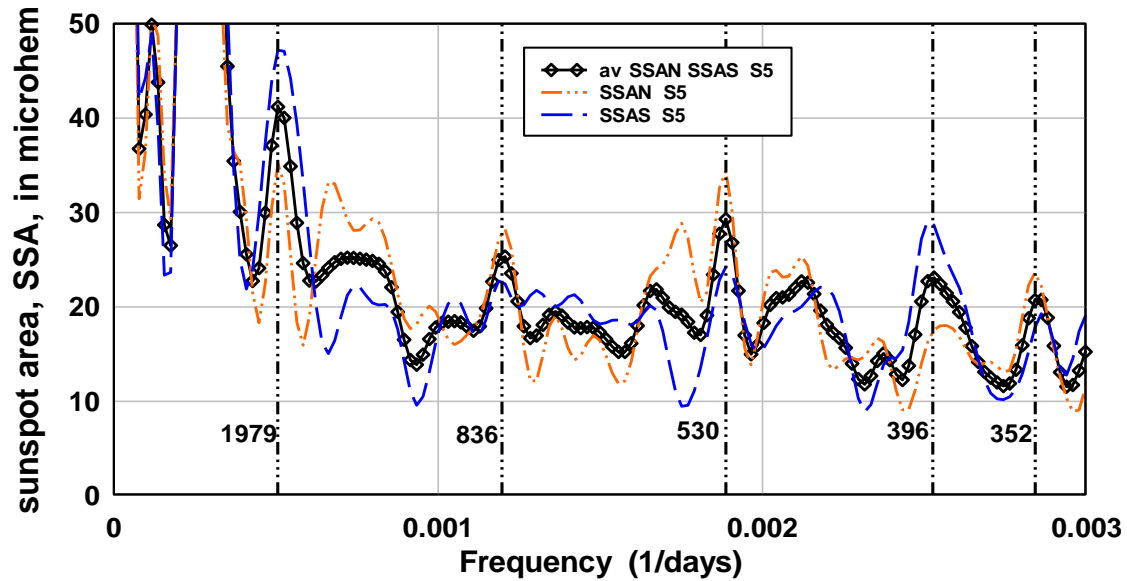

**Figure 1. Spectral content of sunspot area north and south and an average of the two spectra. Each spectrum has been smoothed by a 5 point running average (S5). Spectral peaks common to both the north and south spectrums are indicated with vertical reference lines. The 1876-2012, 50,038 day long, data series allows an accurate estimate of periodicity. The period of each marked peak is an integer number of $T_M/2$ where $T_M$ is the orbital period of Mercury.**

A feature of the marked periods is that all are near exact multiples of $T_M/2$ where $T_M$ is the period of Mercury, 87.96926 days. For example, $352 = (8.003)T_M/2$, $530 = (12.05)T_M/2$, $836 = (19.007)T_M/2$ and $1979 = (44.993)T_M/2$. This suggests that motion of the planet Mercury may be influential in generating periodic sunspot emergence.

Figure 2 shows the periodogram of sunspot area for the period range from 100 days to 10,000 days. In addition to the peaks marked in Figure 1, Figure 2 includes reference lines marking prominent peaks in the period range between 100 and 300 days. Although the latter periodicities occur frequently in spectral analysis of solar activity (Tan & Chen 2013; Kilcik et al. 2010) none have an integer relationship with $T_M/2$ or with any planet orbital period. The broad, strong, peak at around 4000 days is due to the ~ 11 year Schwabe solar cycle, the strongest quasi-periodic feature of all solar magnetic indices.

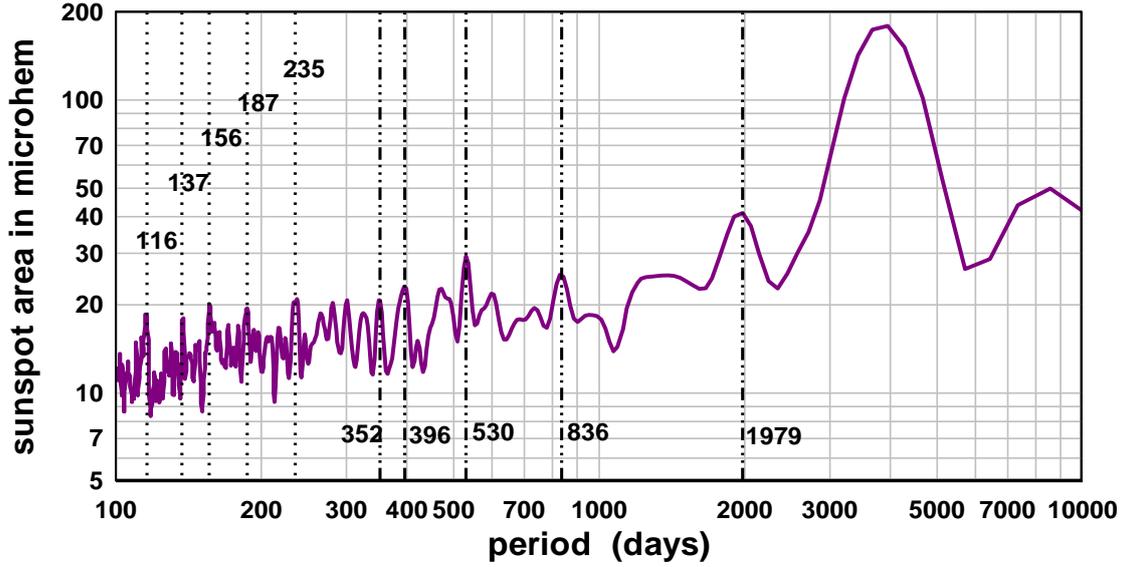

**Figure 2.** Periodogram of average SSAN and SSAS indicating, in addition to the periods marked in Figure 1, the periods of prominent peaks in the 100 – 300 day range. None of the latter peaks have periods with any simple integer relationship to planet orbital periods.

## 4 A PLANETARY MODEL OF SUNSPOT EMERGENCE

Noting the apparent influence of Mercury on periodicity, Figure 1, a rudimentary model of sunspot emergence was developed based on conjunctions between Mercury and the other tidal planets, Venus, Earth and Jupiter. A planetary conjunction corresponds to the alignment of two or more planets on a line with the Sun. This may occur with the planets on the same side or on opposite sides of the Sun. At conjunction a peak in the gravitational tide on the Sun occurs. The basic assumption of the present model is that, by some unspecified mechanism, tidal peaks trigger the emergence of sunspots.

The period of conjunction between planet Mercury and a planet P is given by:

$$T_{MP} = 0.5/(1/T_M - 1/T_P). \tag{1}$$

From the known orbital periods of Mercury, Venus, Earth and Jupiter we find $T_{MV}$ = 72.28311 days, $T_{ME}$ = 57.93874 days and $T_{MJ}$ = 44.89615 days.

In calculating the time variation of conjunctions we use an approach similar to that adopted by Hung (2008). Hung calculated a time dependent alignment index, $I_{MP}(t) = |\cos(\theta_M(t) - \theta_P(t))|$, as a measure of the proximity to conjunction between Mercury and planet P. Here $\theta_P(t)$ is the heliocentric longitude of a planet and the symbols | … | imply the absolute value of the term enclosed. The cosine term takes the maximum value +1 when $\theta_M(t) - \theta_P(t) = 0°$ and Mercury and planet P are aligned on the same side of the Sun and the value -1 when $\theta_M(t) - \theta_P(t) = 180°$ and the planets align on opposite sides of the Sun. Taking the absolute value results in $I_{MP}(t)$ having one maximum during each period, $T_{MP}$. By adding the $I_{MP}(t)$ terms for Venus, Earth and Jupiter one obtains a time varying function that peaks each time one or more of the planets are in conjunction with Mercury. To further simplify the model we give each $I_{MP}(t)$ term equal weight, in the model unity, so that the alignment of Venus, Earth, and Jupiter with Mercury would contribute the quantity 3 to the function.

However, the model developed here differs from that of Hung (2008) by including quadrature terms in the alignment index. The argument for adopting this approach is that when two or more planets are in quadrature to Mercury, i.e. at 90° or 270° to the heliocentric longitude of Mercury, the planets are themselves in alignment with the Sun and a tidal peak occurs. Thus, the alignment indices in the present model are calculated so as to have peaks when $\theta_M(t) - \theta_P(t) = 0, 90, 180$ or 270 degrees.

As the orbits of each planet are elliptical the exact angular difference, $\theta_M(t) - \theta_P(t)$, is a very complex function of time and is usually found by reference to tables of planetary longitude. However, to further simplify the present model and the calculation of $\theta_M(t) - \theta_P(t)$ we assume the planet orbits are circular. In this case the time dependant longitudinal angular difference term can be approximated by a simple sinusoidal function in time. Thus the three individual terms of the quadrature alignment index used in the model become:

$$I_{MV}(t) = |(\cos(2\pi t/T_{MV} + 1.510))|. \tag{2a}$$

$$I_{ME}(t) = |(\cos(2\pi t/T_{ME} + 1.858))|. \tag{2b}$$

$$I_{MJ}(t) = |(\cos(2\pi t/T_{MJ} - 2.1817))|. \tag{2c}$$

Time $t = 0$ is referenced to January 01, 1959, the start of the NASA COHOWEB tabulation of planet coordinates. Although a phase term is not necessary for determining the spectral content of the model a phase term is included so that comparison of the model time variation with the observed time variation of solar activity is possible. Also note that in taking the absolute value of the cosine term the period of each alignment index is halved. Thus, for example, the period of $I_{MV}(t)$ is $T_{MV}/2$, corresponding to the inclusion of two quadrature alignments as well as the usual two conjunction alignments.

Mercury has a highly elliptical orbit. As a result the tidal effect of Mercury, which depends on the inverse cube of the orbital radius, $1/R_M^3$, Hung (2007), varies over a 3:1 ratio with period $T_M$. Taking this variation in Mercury tidal effect into account so as to emphasise the influence of Mercury, we amplitude modulate the alignment index with a term:

$$E_M(t) = 2 + \cos(2\pi t/T_M + 1.563). \tag{3}$$

Thus, the modelled time dependent variation in planetary tidal effect becomes:

$$Y(t) = E_M(t)[I_{MV}(t) + I_{ME}(t) + I_{MJ}(t)]. \tag{4}$$

This variation, of peak value 9, is assumed to be proportional to the combined planetary tide at a point on the solar equator. The variation is plotted in Figure 3a for the first 3000 days from January 01, 1959. The variation has, as expected, a strong component at the orbital period of Mercury, $T_M$, ~ 88 days, and a few shorter period components. There are no components in $Y(t)$ with period longer than $T_M$.

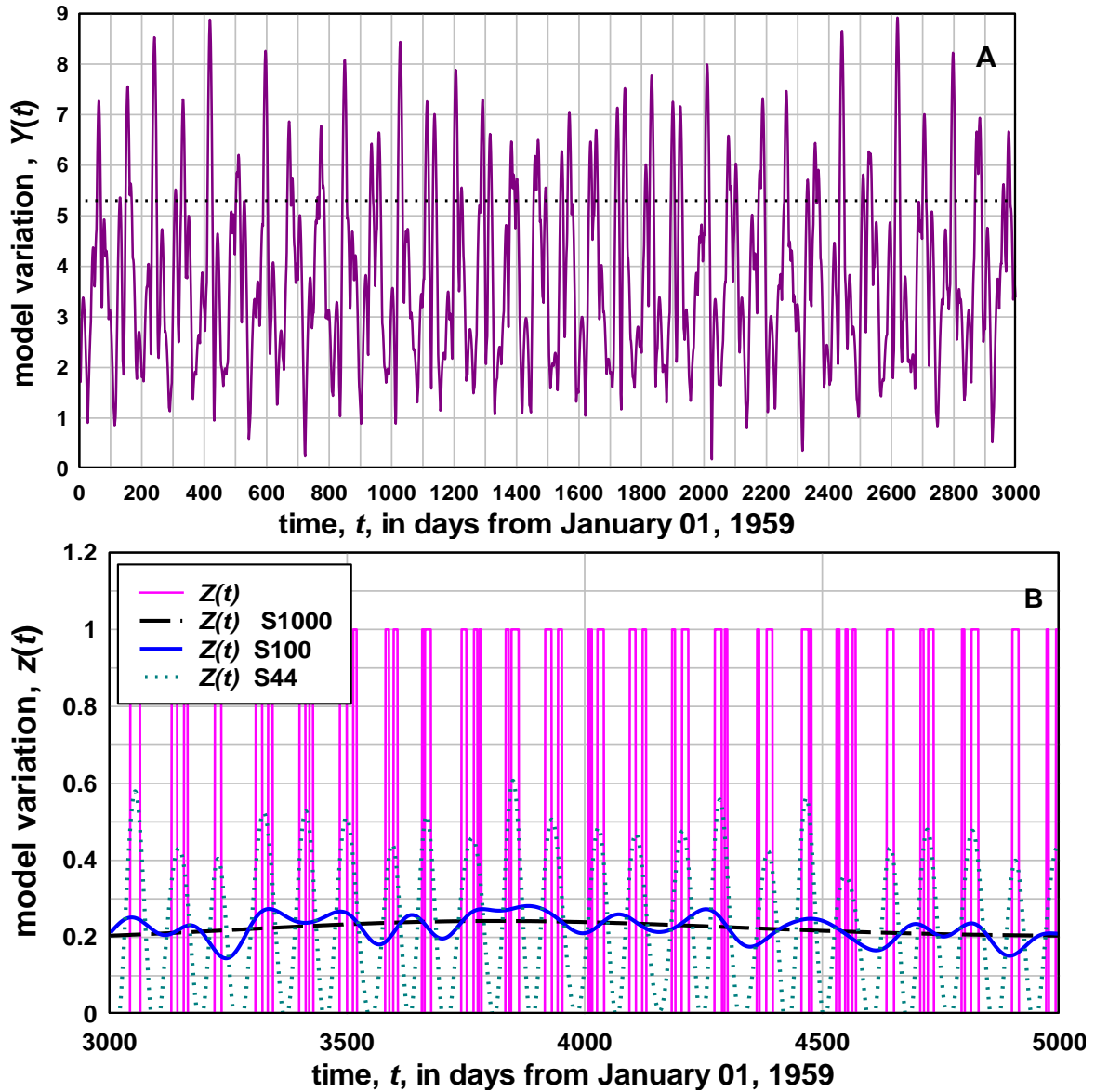

**Figure 3.** (A), the variation of the relative tidal effect, $Y(t)$, at the solar equator as a function of time in days from January 01, 1959. The maximum, value 9, occurs when Venus, Earth and Jupiter are in conjunction or quadrature with Mercury and Mercury is at closest approach to the Sun. The dotted reference line represents an adjustable level above which the model assumes the tidal effect is strong enough to trigger the emergence of sunspots that leads to the modelled time variation of sunspot emergence, $Z(t)$. (B), the function $Z(t)$ changes from 0 to 1 at times when the relative tidal effect, $Y(t)$, exceeds a trigger level and sunspots emerge on the surface of the Sun. Clearly, $Z(t)$ has a strong component at ~ 88 days and at harmonics of the ~88 day component. However, smoothing $Z(t)$ with running averages indicates that $Z(t)$ also has significant components in the intermediate range of periodicity, > 88 days.

To convert the time variation in planetary tide, $Y(t)$, to a time variation in sunspot area we assume that tidal peaks of sufficient amplitude can trigger loops of magnetic flux to float up to the surface to emerge as sunspots. Thus, the single adjustable parameter in the model is a trigger level for sunspot emergence. When the relative tidal amplitude, $Y(t)$, is greater than the trigger level a sunspot is triggered to emerge and the modelled variation of sunspot emergence, $Z(t)$, is given a value unity at time t. If $Y(t)$ is less than the trigger level no sunspot emerges and the modelled sunspot area takes value zero at time $t$. The trigger level applied to the tidal effect, $Y(t)$, is adjusted until the frequency spectrum of $Z(t)$ replicates as closely as possible the observed intermediate frequency spectrum of sunspot area. In the present case a trigger level of relative value 5.3 gave best fit of the $Z(t)$ spectrum to the observed sunspot area spectrum in the intermediate range. This trigger level is shown as the horizontal reference line in Figure 3A. The time variation of $Z(t)$ was calculated from $Y(t)$ in an Excel spreadsheet using the algorithm:

$$Z(t) = \text{if}(Y(t) > 5.3, 1, 0). \tag{5}$$

An example of the resultant variation in $Z(t)$ is shown in Figure 3B.

As $Z(t)$ is a sequence of pulses of variable width the model spectrum has a strong 88 day component and strong components at harmonics of the 88 day component, i.e. components at 44, 29.3, 22, … 88/n days. Significant components at these periods can be found in the spectra of sunspot area or flares, e.g. Oliver and Ballester (1995), but not at strengths consistent with the relative strengths of the 88 day component and the longer period components of $Z(t)$ indicated in Figure 3B. It is suggested that periodicities of 88 days and shorter are attenuated due to the longevity of sunspots and sunspot areas, typically 10 – 100 days, which would tend to smooth out the 88 day and shorter period variations in sunspot emergence. For simplicity, the effect of this attenuation is not included in the planetary model. As indicated in Figure 3B, the $Z(t)$ spectrum also has a wide range of components at periods longer than 88 days. For example, a long period component, of ~ 2500 day period, is evident in the variation of the 1000 day running average of $Z(t)$, Figure 3B. A component of about 150 day period, obtained by a 100 day running average, is also evident. The longer range periodicities appear to arise from variation in the width of the pulses or groups of pulses in $Z(t)$. The components of $Z(t)$ with periods longer than 88 days are in the intermediate periodicity range of interest in this paper.

## 5 MODELLED AND OBSERVED INTERMEDIATE PERIODICITY

The planetary functions $Y(t)$ and $Z(t)$ are chaotic in the sense that the functions do not repeat in the time frame of interest in this paper, < 100 years. Therefore, we expect the spectral content of the model of sunspot emergence, $Z(t)$, to vary or be intermittent over shorter intervals. To illustrate typical model periodicity and intermittency Figure 4 shows the spectra found by taking a FFT over $Z(t)$ in sequential intervals of 16,500 days starting at January 01, 1959. Each 16,500 day interval is approximately the duration of four solar cycles.

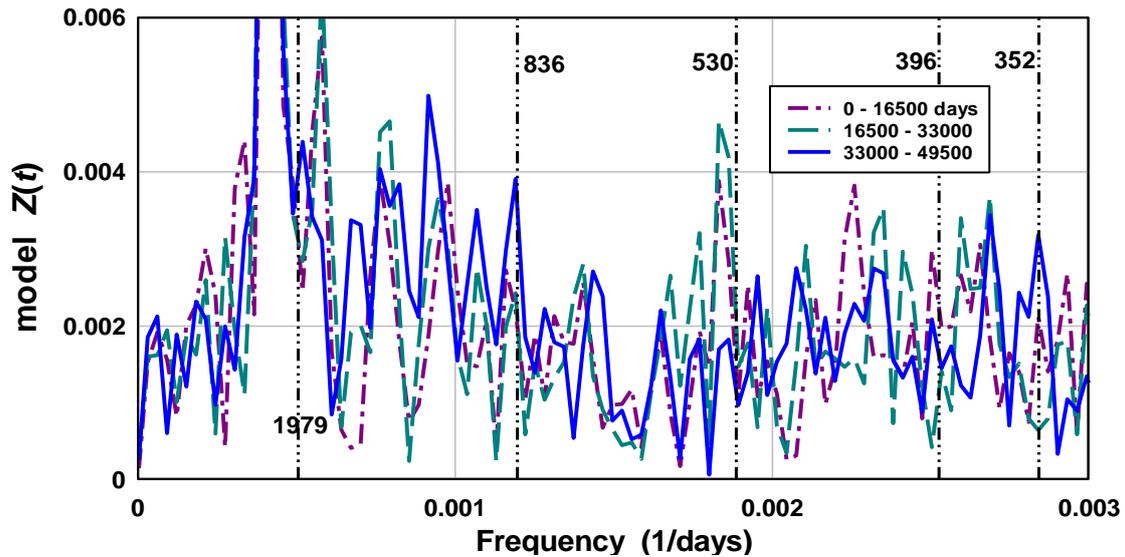

**Figure 4.** When frequency spectra are obtained by FFT over sequential 16,500 day intervals of the sunspot emergence model, $Z(t)$, prominent peaks that are close in period to the prominent peaks marked in the observed sunspot area spectra, Figure 1, are evident. The peaks marked in Figure 1 are marked, at the same period, by reference lines in this figure. Also evident is the intermittent occurrence of spectral peaks from interval to interval.

Evidently, the model intermittently generates spectral peaks some of which are close in period to peaks in the observed sunspot area spectra that were marked in Figure 1. The exception is the 530 day peak which, in the model spectra, appears at 540 days rather than at 530 days. The intermittency of spectral components is evident with, for example, a 352 day peak evident in the spectra from the first and third intervals but absent in the spectrum of the second interval. The model also exhibits numerous other periodicities, some of which may be associated with less prominent peaks in the sunspot area spectra, Figure 1.

When model spectra are calculated over much longer intervals the intermittency is reduced and the spectral components are more clearly defined, Figure 5A.

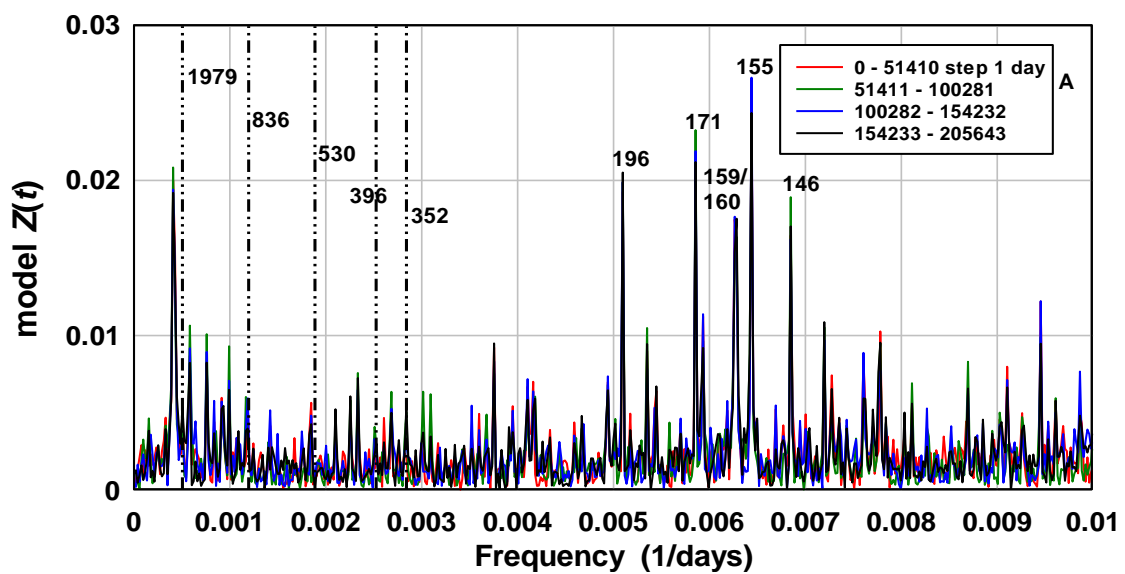

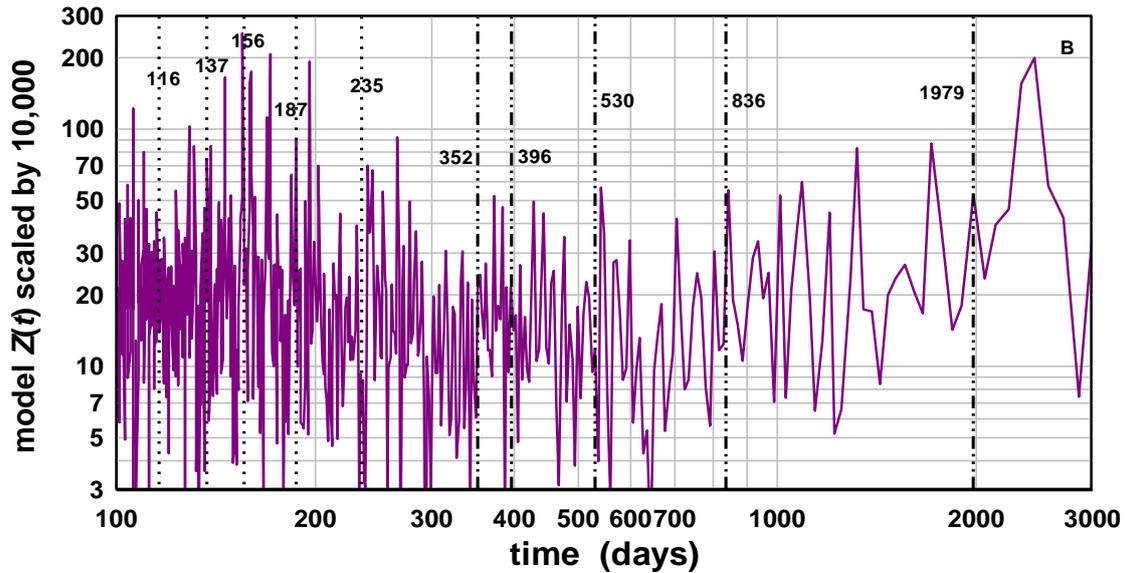

**Figure 5.** (A), when spectra of the planetary model, $Z(t)$, are calculated over sequential intervals of 51,410 days the intermittency is reduced and spectral peaks are well defined. Some of the prominent peaks correspond closely with prominent peaks in the sunspot area spectrum, as indicated by reference lines at the same periods as marked by reference lines in Figure 1. The peak at ~155 days exceeds in strength all other peaks in the model spectrum. The four strong peaks near the ~155 day peak occur at 146, 159/160, 171 and 196 days. (B), the periodogram of the planetary model, $Z(t)$, obtained over the 0 to 51,410 day interval exhibits numerous periodicities in the intermediate range. Here all the periods marked by reference lines in Figure 2 are reproduced by reference lines in this figure.

A noticeable feature of the model spectrum is that the peak at 155 days exceeds all other peaks in strength. Converting the model frequency spectra of Figure 5A to a model periodogram, Figure 5B, allows a comparison with the sunspot area periodogram of Figure 2. The reference lines at periods marked in Figure 2 are reproduced in Figure 5B. All of the components of sunspot area marked in Figure 2 match significant model components apart from the 530 day peak, which in the model occurs at 540 days, and the 235 day peak, which in the model occurs at 240 days. Thus, all ten of the marked sunspot area peaks of Figure 2 are matched within 2% by significant model peaks in Figure 5B. The seven strong model peaks, at periods longer than 1000 days, occur at approximately 2460, 1979, 1713, 1318, 1195, 1094 and 1008 days.

The planetary model generates a large number of peaks in the intermediate range as illustrated in Figure 5. However, it is clear from the number of peaks present in the individual SSAN and SSAS spectra in Figure 1 and in the sunspot area periodogram of Figure 2 that a large number of component peaks is a feature of solar activity spectra when the spectra are obtained over intervals long enough to allow intermittent periodicities to accumulate in the spectrum and long enough to provide sufficient resolution to distinguish individual peaks. It follows that the model replicates two characteristics of intermediate periodicity in solar activity, (1), numerous components and (2), intermittency of components, (Lean 1990, Tan and Chen 2013, Choudhary et al 2014).

An interesting feature of the model spectrum, Figure 5, is that the strongest periodicities occur near 155 days. Closer examination of the spectrum indicates that the strongest peak is at 155.3 days. The periods of the six strong peaks in this vicinity are 146, 155, 159/160, 171 and 196 days. The amplitudes of the model peaks near 155 days are three to five times stronger than the model peaks in the longer 200 day to 1000 day range. This contrasts with the situation in the observed sunspot area periodogram, Figure 2, where the ~ 155 day peak is about the same amplitude as the peaks in the 200 to 1000 day range. Possible reasons for this difference are discussed later.

## 6 OBSERVED AND MODELLED PERIODICITY IN FLARES

The observation by Rieger et al (1984) of ~ 155 day periodicity in hard flares initiated the study of intermediate range periodicity in solar activity generally and, subsequently, there have been numerous reports of ~ 155 day periodicity in flares, (Bogart and Bai, 1985, Ichimoto et al 1985, Bai and Sturrock 1987, Bai and Cliver 1990, Lou et al 2003, Dimitropoulou et al 2008). Thus, strong ~ 155 day periodicity appears to be characteristic of periodicity in flares. While the previous section shows that the spectrum of the model variation, $Z(t)$, provides a reasonably good fit to the sunspot area spectrum we now assess whether the model spectrum provides a good fit to the intermediate periodicity spectrum in flares. Figure 6A shows the spectra of the NOAA north and south flare indices, recorded daily from 1966 to 2008, an interval covering solar cycles 20, 21, 22 and 23. The periods of peaks common to both the north and south spectra are indicated by reference lines. The most prominent common peaks occur at ~542, 204, 158, 148 and 115 days.

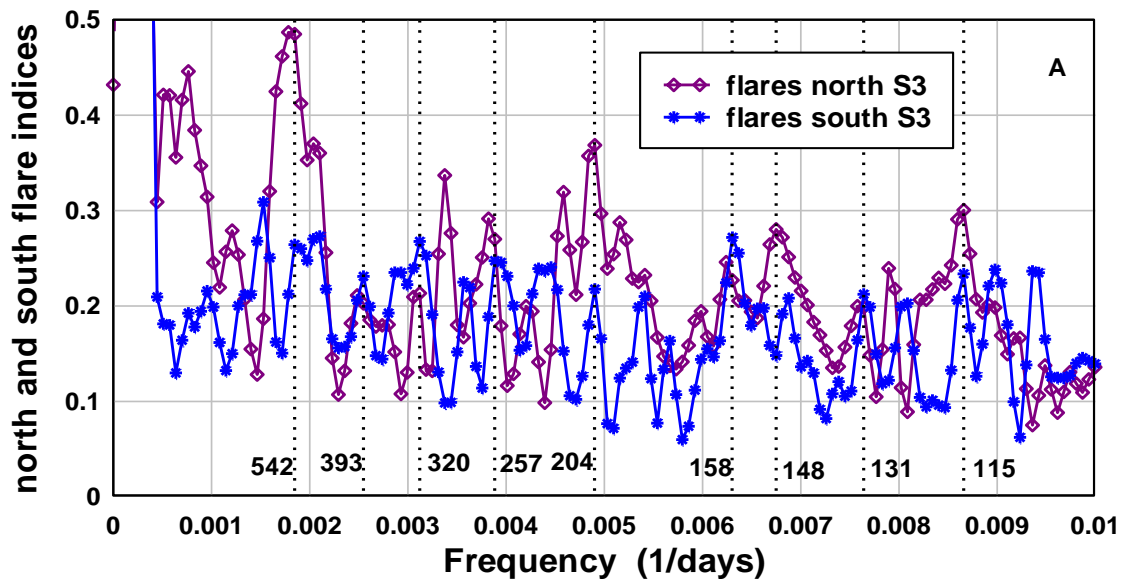

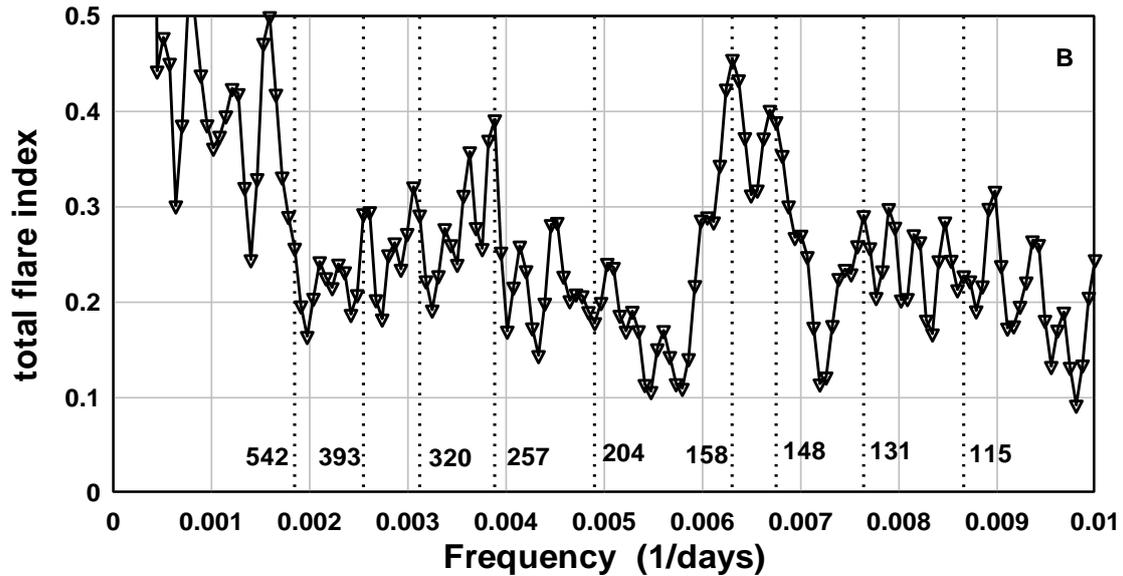

**Figure 6. (A), spectra the daily NOAA flare index, north and south, 1966 to 2008. Periodicities common to the north and south spectra are marked with reference lines and periods in days.
(B), the spectrum of the total flare index. The same periodicities marked in A are reproduced in B. The strength of the 158 and 148 day peaks in B indicates that the north and south components at 158 and 148 days are predominantly in-phase during the interval 1966 to 2008. It is evident, from the absence of corresponding peaks in B, that the 542, 204 and 115 day north and south components are predominantly in anti-phase during the same interval.**

Figure 6B shows the spectrum of the total flare index with the same references lines and marked periods as in Figure 6A. It is evident that ~155 day periodicity, corresponding to the frequency range 0.006 to 0.007 days$^{-1}$, (167 – 143 days), provides a very significant fraction of power in the intermediate range of periodicity in the NOAA total flare index. This suggests that the components of the ~ 155 day periodicity on the north and south hemispheres are in-phase during most of the 1966 to 2008 interval. Other in-phase components appear to be at ~ 257 and ~ 393 day period. Similarly, it is evident that the ~ 542, ~204 and ~115 day components are predominantly in anti-phase during this 1966- 2008 interval. These results suggest that the ~ 155, ~257 and ~393 day narrowband components of the total flare index could be usefully compared with the corresponding narrowband components of the planetary model to assess predictive capability of the model.

Figure 7 compares the spectrum of the NOAA total flare record with the model spectrum in the ~ 155 day period range. Two of the strongest peaks in the flare spectrum are at 155 and 160 days, and correspond closely in frequency with the 155 and 159/160 model peaks in this range.

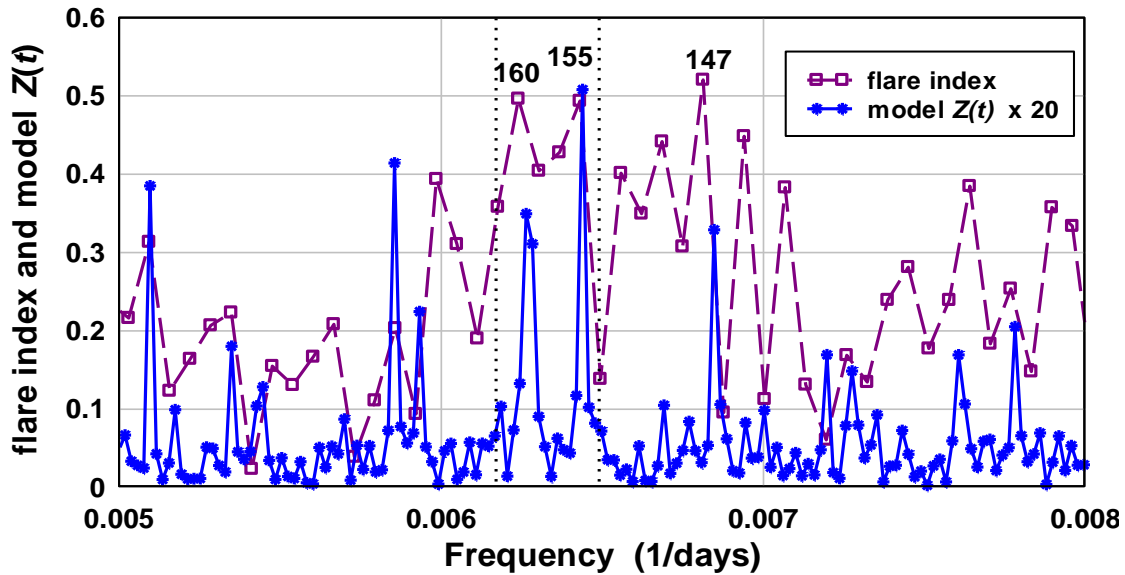

**Figure 7.** The planetary model, $Z(t)$, spectrum scaled for comparison with the total flare index spectrum in the frequency range centred on ~155 day periodicity. The correspondence of the spectra in the frequency region between the dotted reference lines suggests that the components there could be used to generate narrow band variations of the flare index and the model for comparison of ~ 155 day periodicity in the time domain.

In summary, the planetary model, $Z(t)$, has many components, a significant number of which occur at periods close to the periods of components in the flare index. Near 155 day period components dominate in the model spectrum and in the observed total flare spectrum. This suggests it may be feasible to use the components in this range to generate the ~155 day period time variations of the model and the total flare index for comparison and assessing the possibility of prediction using the model.

## 7 CAN THE MODEL PREDICT THE ~155 DAY FLARE COMPONENT

In principle, a planetary model of solar activity, based on the predictable motion of the planets, should be able to predict the time variation of the various components in solar activity, for example the time variation of the ~155 day component of flare activity. Referring to Figure 5A, we note that the model variation has its strongest component at 155 days. This single, 155 day period, component could provide a constant amplitude sine wave that could be compared, in phase, with the observed time variation of the ~155 day component of flare activity. However, a single frequency component would provide no information in respect to the intermittency that seems to be a characteristic of ~155 day flare activity. For example, a single component variation would provide no information on why the ~ 155 day periodicity is strong in solar cycle 21 and weak in solar cycle 22, (Kile & Cliver 1991; Bai 1992; Oliver & Ballester 1995). Predicting intermittency requires selecting a frequency band that contains more than one frequency component of the model so that interference between the components generates a signal that varies in amplitude.

Referring to the spectra in Figure 7, we select the frequency range between the dotted reference lines to reproduce narrow band flare and narrow band model variations for comparison. This range includes four flare index spectral points and three strong model spectral points. Summing the time variations of the components corresponding to these points generates ~ 155 day period variations of the flare index and the model. The co-sinusoidal contributions of each selected point, obtained from the FFTs, are as follows:

Total Flare Index, reference time t = 0 on January 01, 1966:
$F_1(t) = 0.496\cos(2\pi 0.00623965 t + 2.7668).$ (6a)
$F_2(t) = 0.404\cos(2\pi 0.00630332 t - 0.58732).$ (6b)
$F_3(t) = 0.427\cos(2\pi 0.00636699 t - 2.64335).$ (6c)
$F_4(t) = 0.494\cos(2\pi 0.00643066 t + 1.1872).$ (6d)

Model, reference time t = 0 on January 01, 1959:
$Z_1(t) = 0.0254\cos(2\pi 0.006438436 t - 2.204289).$ (7a)
$Z_2(t) = 0.0175\cos(2\pi 0.006263373 t + 1.792887).$ (7b)
$Z_3(t) = 0.0155\cos(2\pi 0.006282824 t - 1.423750).$ (7c)

As a preliminary, we compare the narrow band model variation with the raw flare index variation. After taking into account the reference time difference of 2557 days, the three point narrowband model variation, scaled by 200, is shown in Figure 8 along with the total flare index and a 78 day running average of the flare index. As expected, the three component planetary model varies significantly in amplitude indicating it may be useful in predicting when ~155 day periodicity is likely to be weak or strong. For example, the model variation predicts that ~155 day periodicity is weak in solar cycle 22, consistent with observations, (Kile and Cliver 1991, Bai 1992 and Oliver & Ballester 1995), strong in solar cycles 20, 21, 23, (Dimitropoulou et al 2008), and, on the basis of the extended model variation, ~155 day periodicity is predicted to be strong in solar cycle 24.

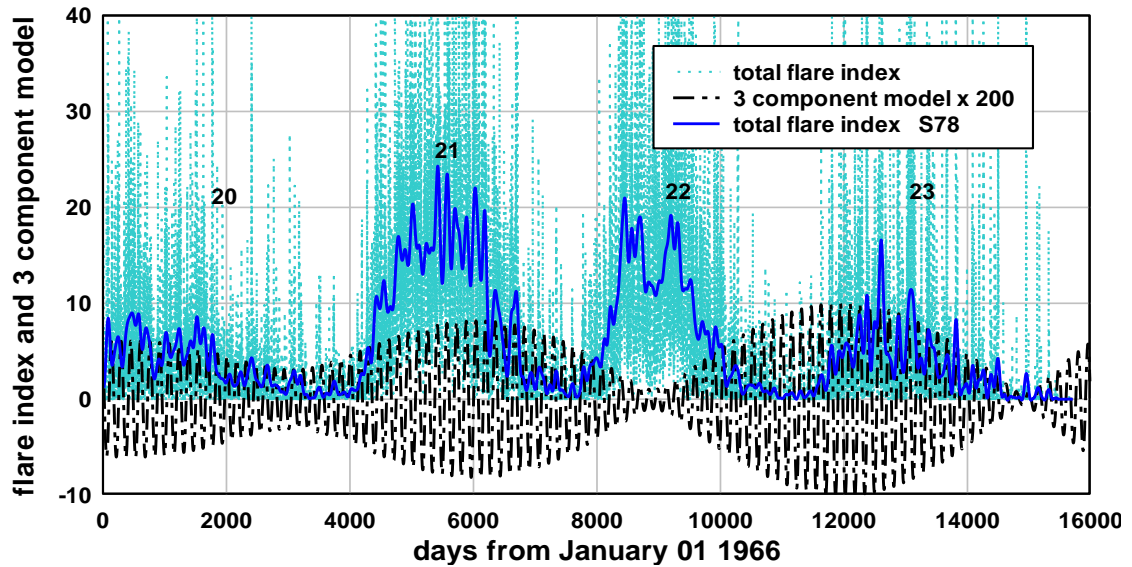

**Figure 8. Scaled comparison of the flare index, the 78 day running average of the NOAA flare index and the narrow band, ~155 day period, variation of the planetary model. With three components contributing to the model, the amplitude varies significantly and predicts when the ~155 day component in solar activity should be weak or strong.**

We now compare the narrow band flare and model variations. The 3 component, ~ ~155 day, model variation, scaled to match, is compared with the 4 point, ~ 155 day, flare variation in Figure 9. A noticeable feature is that the long term amplitude variation of the ~155 day model is similar, but not exactly the same, as the long term amplitude variation of the ~155 day flare component. This supports the idea that the narrow band approach may have longer term predictive capability as to when ~155 day periodicity is expected to be strong. Note that the narrow band model contains no components relating to the ~ 11 year Schwabe cycle and can only indicate the long term amplitude variation of the ~155 day cycle in the model. The short term capability to predict when peaks in the ~ 155 day flare cycle occur within a solar cycle appears to be excellent. Figures 10A and 10B expand the time scale of Figure 9 to illustrate the comparison of the short term ~155 day variations of the model and flares in more detail.

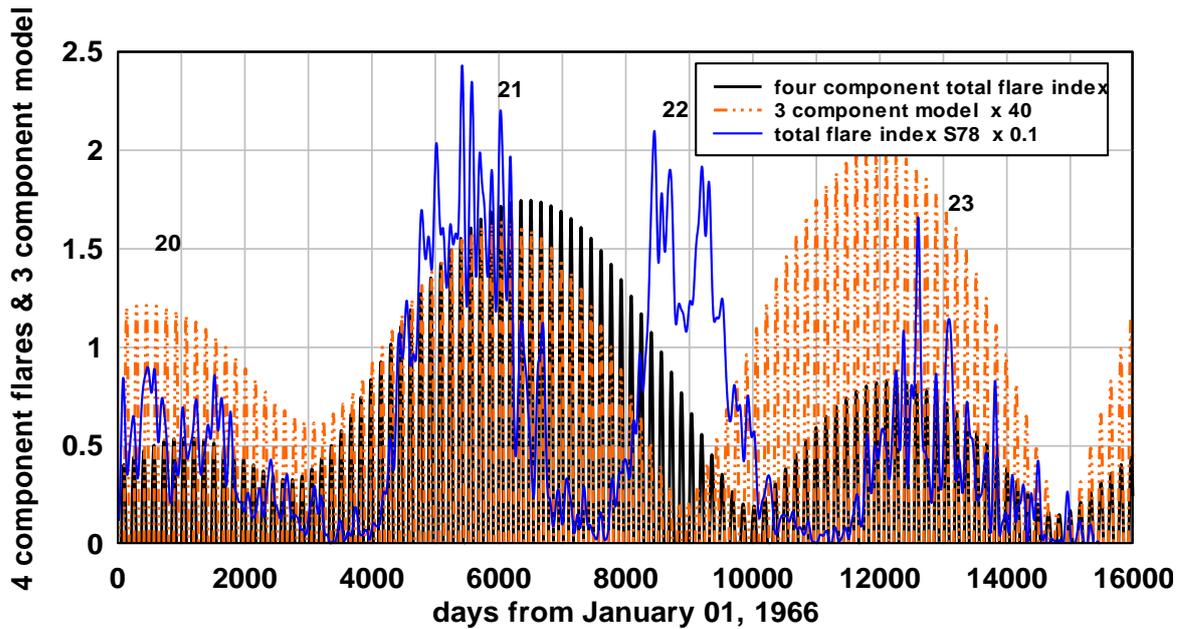

**Figure 9. A comparison of the 4 component, ~ 155 day, variation of the total flare index with the 3 component, ~ 155 day, variation of the planetary model, $Z(t)$, over the entire interval of the NOAA flare index. A 78 day running average of the flare index, scaled for comparison, is also shown. The longer term amplitude variations of the two narrowband variations are similar with both showing sharp minima during solar cycle 22 and near the end of solar cycle 23.**

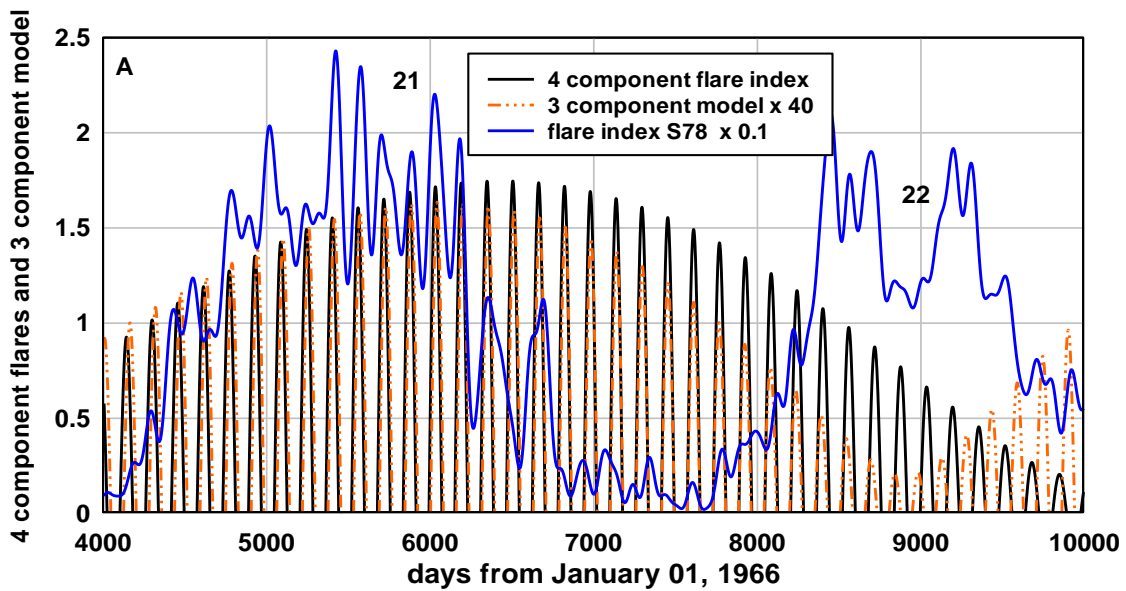

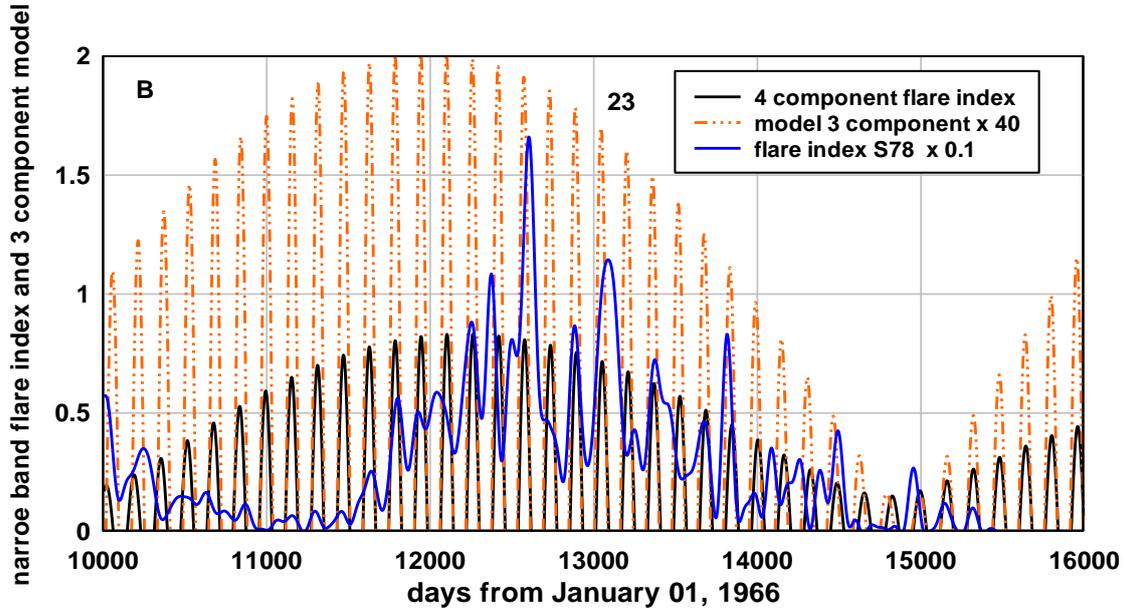

**Figure 10.** (A), a comparison of the narrowband, ~ 155 day period, variations of the NOAA total flare index and the planetary model, $Z(t)$, during solar cycles 21 and 22. A scaled version of the 78 day running average of the flare index is included. During solar cycle 21 there is close correspondence between the peaks in the model narrowband variation and peaks in the flare variation. (B), compares the narrowband, ~ 155 day period, variations of the NOAA flare index and the planetary model, $Z(t)$, during solar cycle 23. During solar cycle 23 there is a very close correspondence between the observed narrowband and modelled narrowband variations.

The correlation coefficient between the two narrow band signals in the range 0 – 4,000 days, (solar cycle 20) is 0.25; in the range 4,000 – 10,000 days, (solar cycles 21 and 22), is 0.84; and in the range 10,000 to 15,706 days, (solar cycle 23), is 0.94. The correlation coefficient over the entire 15,706 day range is 0.73. It is worth noting that this high correlation is obtained between two narrowband signals, one derived from planetary motion and the other from flare observations. Except for the short interval, 9,000 – 10,000 days in Figure 9 and Figure 10A, at one of the long term minimums of the model variation, when the ~155 day model and flare variations differ in phase by up to $\pi$ radians, the two narrowband variations are in-phase and this coherency is maintained over solar cycles 21 – 23. This is consistent with observations by Lean (1990) who showed that ~155 day periodicity in solar activity remains approximately coherent between solar cycles.

The overall pattern of the 3 component, ~ 155 day period, planetary model repeats every ~52,000 days, Figure 11, with an intermediate pattern that repeats every ~6,000 days. As the solar cycle, indicated for the latest five solar cycles in Figure 11, repeats every ~4000 days it follows that every third solar cycle maximum should, on average, coincide with one of the sharp minima in the longer term model variation. For example, as solar cycle 22 coincides with a model minimum it can be predicted that solar cycle 25 should coincide with the next model minimum and solar cycle 28 should coincide with the following model minimum. The solar cycles in between, e.g. solar cycles 23 and 24 and solar cycles 26 and 27, would be expected to show strong ~ 155 day activity. It has been reported that solar cycle 23 shows clear evidence of ~ 155 day periodicity, (Krivova & Solanki 2002; Ballester et al. 2004; Dimitropoulou et al. 2008; Chowdhury et al. 2009; Lobzin et al. 2012; Singh et al. 2018; Zaqarashvili & Gurgenashvili 2018). Chowdhury et al. (2015) report ~155 day periodicity in solar activity in solar cycle 24. However, the Barlyaeva et al. (2018) report that ~155 day periodicity in flares is absent in solar cycles 23 and 24, differs from the observations for solar cycle 23, and the prediction for solar cycle 24 of this paper, Figure 10B.

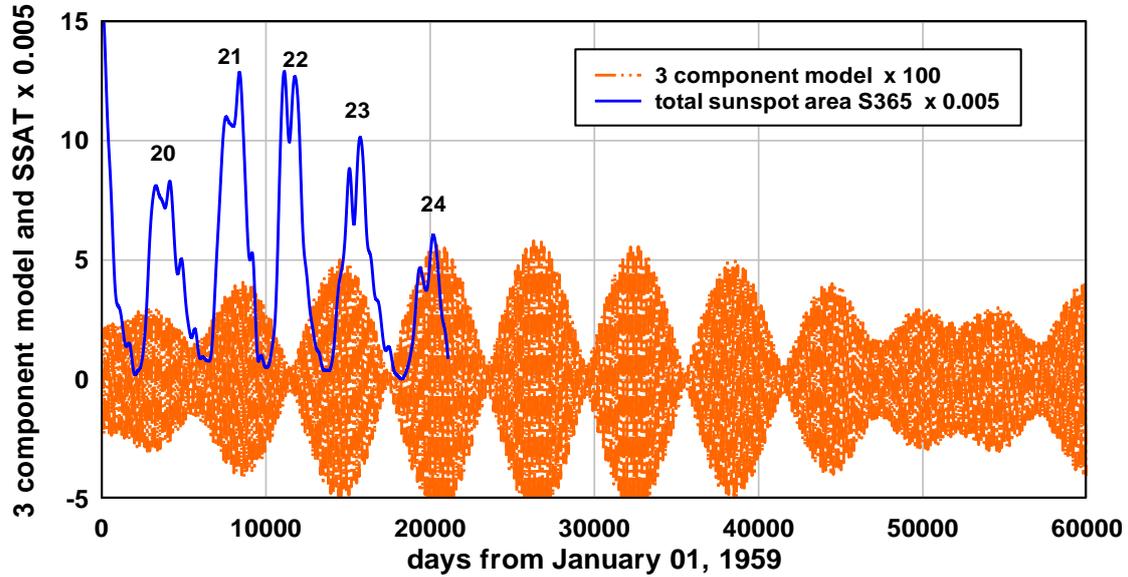

**Figure 11. If the three component narrowband, ~155 day period, variation of the planetary model, $Z(t)$, is extended into the future by about 100 years a broad pattern of maxima and minima is evident. Sharp minima occur at ~6,000 day intervals and longer term minima occur at ~52,000 day intervals. A smoothed and scaled version of total sunspot area is included. The coincidence of one of the ~6000 day period sharp minima with a solar maximum implies very low ~155 day periodicity in that solar cycle.**

We saw, in Figure 9, that the longer term amplitude variation in the ~155 day component of flares correlated moderately well with the longer term amplitude variation of the ~155 day component of the model in that the times of amplitude maxima and amplitude minima of the two variations coincided well. However, the relative amplitudes of the long term maxima differed significantly. For example, Figure 9 shows similar flare and model maximum amplitudes in solar cycle 21 but significantly lower flare maximum amplitude than model maximum amplitude for solar cycles 20 and 23. The simple reason for this appears to be that solar cycles 20 and 23 are significantly weaker than solar cycle 21 and this is reflected in the ratio between the model and the flare maximum amplitudes. A more accurate prediction of the amplitude of the ~155 day variation in flare activity could be obtained if the ~155 day model $Z(t)$ variation was amplitude modulated by the amplitude of the solar cycle, if the amplitude of the solar cycle could also be predicted, as, for example, by methods discussed by Scafetta (2012).

Lean (1990) investigated ~155 day periodicity in sunspot areas for solar cycles 12 to 21 and showed that the amplitude of the periodicity was strongest in solar cycles 18 and 19. The total sunspot area for solar cycles 18 to 20, after smoothing with a 78 day moving average to select for ~155 day periodicity is shown in Figure 12. The ~ 155 day, 3 component model of $Z(t)$ is also shown. The interval spacing of the graph is set to 156 days to facilitate identifying peaks that occur at this periodicity. From the near coincidence of sunspot area peaks with the 156 day reference lines it is evident that ~ 155 day periodicity is coherent over the four solar cycles in the interval. We have previously shown ~ 155 day coherency extends from solar cycle 21 to 23, Figures 9 and 10. Thus, the data in Figure 12 is an indication that coherency of ~ 155 day periodicity extends over at least six solar cycles, 18 to 23. An interesting feature is that small coherent peaks are evident during the solar minima, for example, during the solar minimums between solar cycles 19 and 20 and solar cycles 20 and 21. This suggests that ~155 day periodicity in solar activity is not confined to solar maxima. We note that while, according to Figure 10A, ~155 day periodicity is strong in solar cycle 21, Figure 12 indicates it is also very strong in solar cycle 18. In solar cycle 21 coherent peaks occur regularly, while solar cycle 18 has stronger but less regular occurrence of coherent peaks.

With the 3 component planetary model plotted on Figure 12, we note that, as expected, there is a fairly close coincidence of peaks of the model variation and peaks in sunspot area. We also note that in this time interval the overall amplitude of the planetary model does not fall to the sharp minima in amplitude at ~ 6000 day intervals, as indicated in Figure 9 or 11. So, in the time interval encompassing solar cycles 18,19, 20 and 21, the ~155 day component of the model is strong and of relatively constant amplitude. This is reflected in the strong ~155 day component in sunspot area observed in these solar cycles by Lean (1990) and Carbonell & Ballester (1992) providing further support for predictive capability of the model.

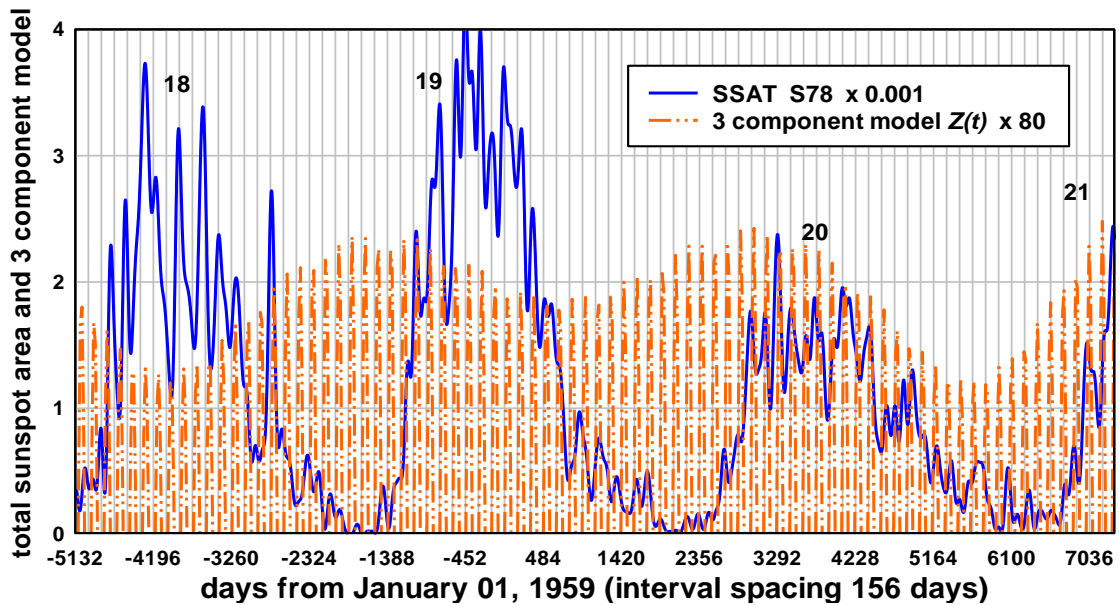

**Figure 12. In this graph the 3 component, narrowband, ~155 day period, variation of the planetary model is projected back from January 01, 1959, day 0 on the graph, to encompass the time interval that includes solar cycles 18 to 21. Comparison of the model variation with the 78 day smoothed variation of sunspot area confirms that the ~155 day model peaks coincide fairly closely with the ~155 day peaks of sunspot area activity over the entire interval. The correspondence of the model variation with the coherent occurrence of ~155 day solar activity during this interval provides further support for the predictive capability of the planetary model.**

## 8   COMPARING NORTH AND SOUTH FLARE ACTIVITY

The premise of the planetary model is that peaks in the ~155 day component of $Z(t)$ trigger sunspot emergence and result in the occurrence of peaks at ~155 day period in sunspot and flare activity. When the components of solar activity on the north and south hemispheres are in-phase it is reasonable, within the premise of this model, to expect that the corresponding planetary model component and the north and south components of solar activity would be in-phase and strongly correlated.  However, usually, short term solar activity on the North and South hemispheres is not well correlated, (Norton & Gallagher 2010; McIntosh et al. 2013). Nevertheless, as indicated in the figures above, the ~155 day components of the total flare variation and model variation are strongly correlated during most of the interval from solar cycles 18 - 23. To study the hemispheric situation more closely, Figure 13 compares smoothed versions of the North and South flare indices during solar cycle 21 when ~ 155 day periodicity was strongly observed in the total flare index. The time scale of Figure 13 is divided into intervals of 155 days to facilitate seeing when peaks at this periodicity occur. The peaks in the ~155 day component of the planetary model $Z(t)$ variation are also shown. The north and south flare index variations are clearly not well correlated. Further, it is clear from Figure 13 that ~155 days periodicity is dominant in southern flare activity while a longer, ~ 500 day periodicity, appears to dominate northern flare activity. However, notice, in Figure 13, that the peaks that do occur in both the northern activity and southern activity are near coincident with the ~ 155 day, planetary model peaks. Also, when a ~155 day peak is absent in the southern activity, a large peak of northern activity occurs, for example at days 4495, 5580 and 6045 in Figure 13. As a result the total flare index for solar cycle 21 is very strongly of ~ 155 day periodicity, as observed, Figure 9 and Figure 10A of this paper, and by Rieger et al (1984) and others.

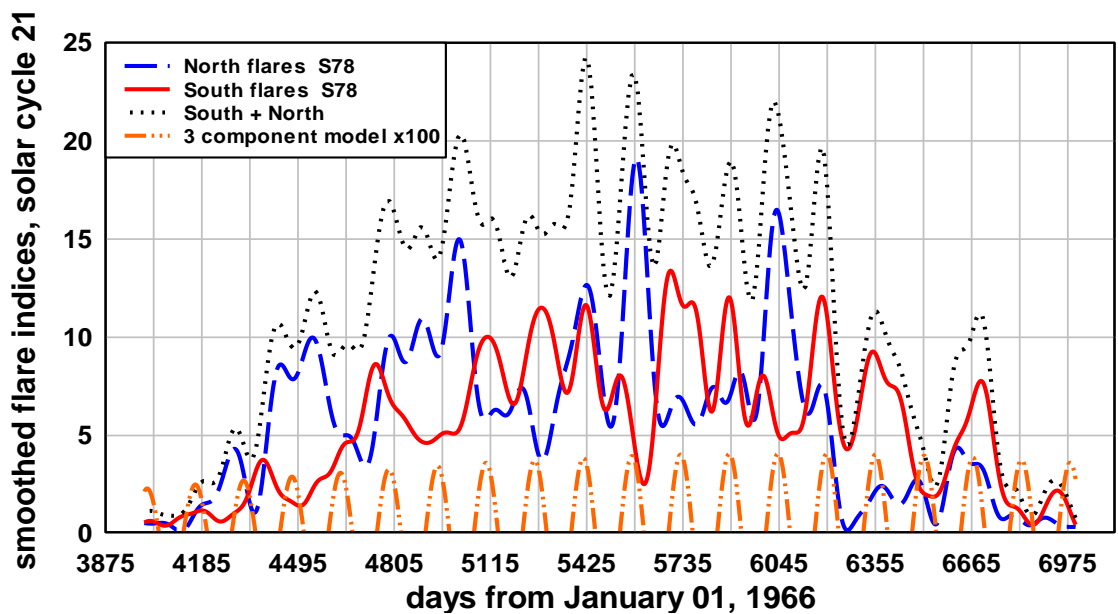

**Figure 13.** This graph shows 78 day running averages of north flares and south flares and total flares during solar cycle 21. Also shown are the positive peaks of the narrowband, 3 component variation of the planetary model $Z(t)$. The time axis is divided into 155 day intervals to facilitate observing 155 day periodicity. It is evident that a ~155 day component dominates in south flares and a ~ 500 day component dominates in north flares. However, the peaks that do occur in both the northern activity and southern activity are near coincident with the ~ 155 day, planetary model peaks suggesting a planetary influence common to both hemispheres. As a result, ~155 day periodicity is very strong in the total flare index activity during solar cycle 21.

We interpret the results in Figure 13 in terms of the planetary model as follows. Peaks in the narrowband, ~155 day component of the planetary model, $Z(t)$, indicate when sub surface magnetic flux is likely to be triggered to float to the surface as a sunspot. However, triggering depends on the sub surface magnetic flux being primed for and close to flotation. When a sunspot does emerge a flare is only likely if the sunspot emerges within an existing sunspot or sunspot group, (Heyvaerts et al 1977, Nishio 1977, Ballester et al 1999, Ballester et al 2002, 2004). Thus, the occurrence of sunspots or flares is not necessarily associated with every peak of the planetary model variation. In Figure 13, it appears that, by chance, on the southern hemisphere nearly every model ~155 day peak is associated with a peak in flare activity while on the northern hemisphere, by chance, every third ~ 155 day peak of the model variation results in a peak in flare activity. However, as indicated above, when, occasionally, a ~ 155 day peak in south flare activity was absent, the absence was compensated by the occurrence of a strong peak in north flare activity. The result is that the ~155 day component of the total flare activity is very strong. It may also be relevant that the northern hemisphere peaks, while about 1/3 less frequent are about 3 times the strength of the southern hemisphere peaks. This suggests that, at this periodicity, about the same amount of magnetic flux is released in sunspots and about the same energy is released in flares on each hemisphere during this solar cycle.

The premise of the planetary model is that when peaks in the periodic variations associated with the model, Figure 3B, occur, unstable subsurface magnetic flux tubes are triggered to float up to the surface. Thus, a delay is expected between the occurrence of a model peak and the emergence of sunspots on the surface and the occurrence of flares. Close examination of Figure 13 indicates that delays between corresponding model peaks and flare peaks vary from 75 days to -5 days. Clearly the time resolution is not adequate to closely define a delay. However, the observed delay range is consistent with theoretical estimates that the rise times of magnetic flux tubes is between 10 and 100 days, Moreno-Insertis (1983).

It follows from the above analysis that the interpretation of solar activity spectra can, in some cases, be difficult. For example, if sunspot activity and consequent flare activity, in response to the ~155 day period planetary triggering, as illustrated in Figure 13, had followed the same time variation on the south hemisphere as on the north hemisphere during solar cycle 21, then Rieger et al. (1984) would have observed ~500 day periodicity rather than ~155 day periodicity in total flare activity.

The same four frequencies as used in the narrowband filter for total flares can be used to obtain narrowband variations of the individual northern and southern flare indices. The components associated with the four points in the FFTs used to generate the narrowband flare variations are:

North flares, reference t = 0 on January 01, 1966:
$N_1(t) = 0.30\cos(2\pi 0.0062396t + 2.321001)$. (8a)
$N_2(t) = 0.17\cos(2\pi 0.0063033t - 1.2627343)$. (8b)
$N_3(t) = 0.20\cos(2\pi 0.0063670t - 2.8747106)$. (8c)
$N_4(t) = 0.27\cos(2\pi 0.0064306t + 1.4472720)$. (8d)

South flares, reference t = 0 on January 01, 1966:
$S_1(t) = 0.26\cos(2\pi 0.0062396t - 2.983519)$. (9a)
$S_2(t) = 0.29\cos(2\pi 0.0063033t - 0.209640)$. (9b)
$S_3(t) = 0.24\cos(2\pi 0.0063670t - 2.445668)$. (9c)
$S_4(t) = 0.24\cos(2\pi 0.0064306t + 0.893496)$. (9d)

While corresponding North and South components in the above relations appear to differ significantly in amplitude and phase angle the component sums are similar in time dependent amplitude variation and are, for most of the 15,706 day interval, close to in-phase and in-phase with the three component model, Figure 14.

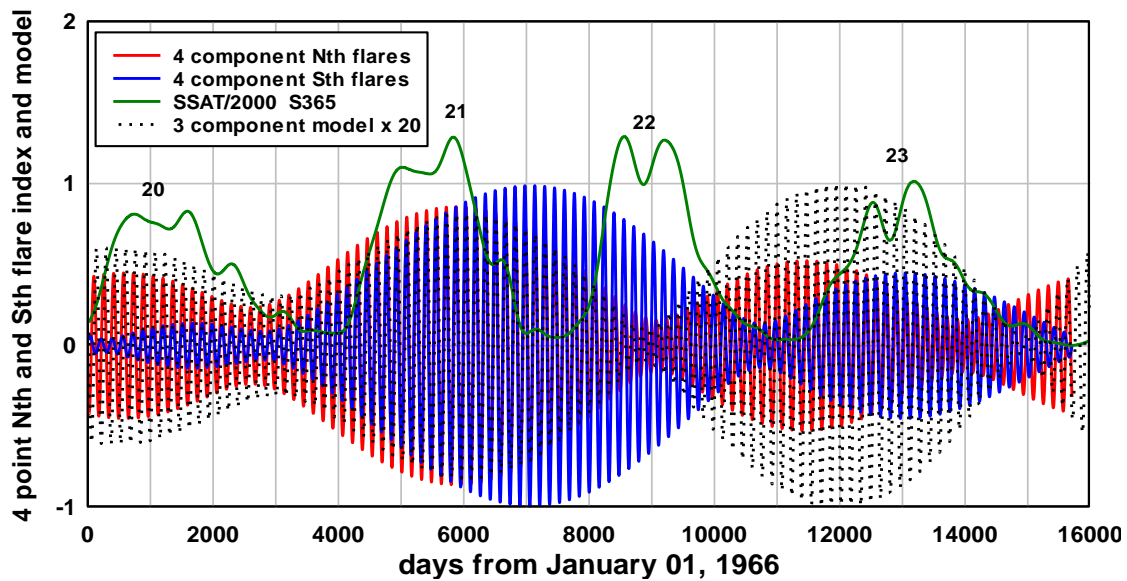

**Figure 14.** The narrowband, ~ 155 day period, variations of the north and south flare indices are, apart from brief intervals around 10,000 days and 15,000 days, close to in-phase and in phase with the ~ 155 day, three component version of the planetary model. This suggests that sunspot activity on the north and south solar hemispheres is, for most of the time, responding in-phase with a planetary influence of ~ 155 day periodicity.

The ~155 day northern flare variation is almost exactly in phase with ~155 day southern flare variation except for short intervals near 10,000 days and near 15,000 days in Figure 14. At these times, during intervals of about 1000 days, a significant phase difference, about π radians, develops between the north and south variations after which the phase difference reverts to zero. These times also coincide with sharp minima in the model variation. Thus, except for the short intervals mentioned, the ~ 155 day components of north flares and south flares are in-phase and in-phase with the 3 component, ~155 day, model variation. During the interval shown in Figure 14 the ~155 day component of north flares is strongly in-phase with the model variation, (correlation coefficient 0.74), and closely approximates the amplitude variation of the ~ 155 day component of the model. The ~ 155 day component of south flares is less strongly in-phase, (correlation coefficient 0.58), and the amplitude is less well matched with model. Nevertheless, the result is consistent with the analysis above which provided support for magnetic activity on both hemispheres being influenced by the ~ 155 day planetary model variation. The result suggests that, although narrowband filtering has the problem of poor time resolution, it can be useful in determining when a particular periodicity is influencing solar activity variation. For example, the narrowband filtering of north and south flare activity illustrated in Figure 14, detected the in-phase presence of ~155 day periodicity in both north and south flare activity during solar cycle 21. However, FFTs of north and south flares during solar cycle 21 detect predominantly ~155 day periodicity in south flares and predominantly ~500 day periodicity in north flares. Thus a simple interpretation of the FFT spectra would be inconsistent with the above evidence that both north activity and south activity are responding, in-phase, to the ~ 155 day planetary model component.

We now consider briefly whether a correlation between flare components and model components at other periodicities can be detected. Figure 6B shows that there is moderately strong periodicity in the flare index at ~ 390 days and Figures 5A and 5B indicate that there is a moderately strong planetary model peak at 384 days which could be driving the similar periodicity in flares. In order to assess whether a planetary influence on flare activity is evident at this periodicity we here correlate simple band pass versions of the time variation of the NOAA flare indices with the single frequency component version of the planetary model, Figure 15. A single frequency component gives no information on the strength of a corresponding flare component but can indicate if there is phase coherence between the two.

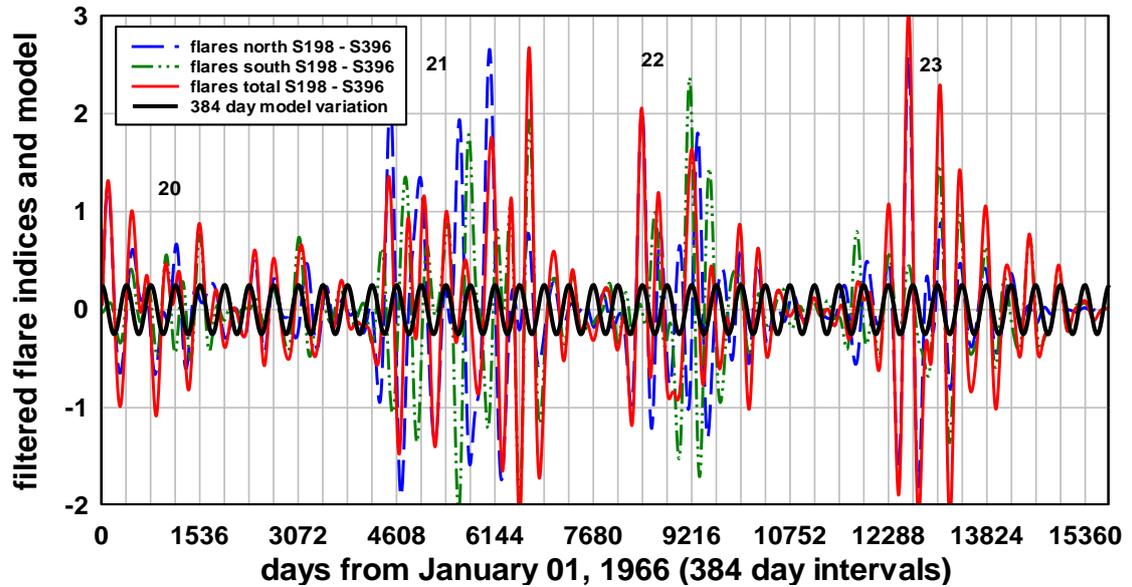

**Figure 15. Simple narrowband versions of the ~388 day components of north, south and total flare indices and the 384 day component of the planetary model, $Z(t)$. It is evident that a close phase relationship with the model variation exists only when the north and south components are in-phase. This occurs mainly in solar cycle 23 and in solar cycle 20. The narrowband versions of the flare indices were calculated as the difference between 198 and 396 day running averages of the indices.**

The simple band-pass versions of the ~ 390 day components in the flare indices were calculated as the difference between a 198 day running average and a 396 day running average. The narrowband north, south and total flare indices are compared with the 384 day component in the model, $Z_{384}(t) = (0.0047)\cos(2\pi 0.0026065t + 1.981)$, after scaling the model to amplitude 0.25. The single frequency model component is moderately in-phase with the total flare component in cycle 20, and more strongly in phase in cycle 23. The correlation coefficient over the four solar cycles is +0.20. Again, correlation between the model component and the total flare component is significantly positive only when the north and south flare components are moving in-phase, as in solar cycle 23. During solar cycle 23 the correlation coefficient between the filtered total flare variation and the model variation is 0.48.

In summary, the results suggest that provided the north and south components of periodic flare activity are in-phase it is likely that the corresponding planetary model component will be in-phase with the north, south, and total components of flare activity

## 9 DISCUSSION AND CONCLUSION

Analysis in the previous sections 7 and 8 indicates that the planetary model has some capability for predicting the occurrence of ~ 155 day Rieger periodicity in sunspot and flare activity. However, prediction is only effective if the north and south components of solar activity are varying in-phase. The result in Figure 14 indicates that north and south, ~155 day periodicity in flares, is in phase in solar cycles 20 and 21, becomes anti-phase in solar cycle 22 and reverts to in-phase in solar cycle 23. Narrowband filtering of sunspot area data from solar cycles 13 to 23, (not shown here), indicates in-phase north and south ~155 day periodicity extends over solar cycles 17 to 21, becoming anti-phase in solar cycle 22 and reverting to in-phase in solar cycle 23. However, north and south ~155 day periodicity in sunspot area is anti-phase in solar cycles 13 and 14. Thus, it is clear that in-phase variation of north - south components of solar activity at ~155 day period does not always occur. Similarly, the results in Figure 6 indicate variation of north and south components at ~540, 200 and 115 day periodicity are mainly anti-phase during solar cycles 20 to 23. There are significant components at these periodicities as well as the strong ~ 155 day period component in the planetary model variation. However, it is difficult to understand how a tidal component derived from the simple planetary model of this paper could be triggering north and south components of activity that are in anti-phase. Nevertheless, hind casting, during solar cycles 20 – 23 when north and south components of flare activity at ~155 day period were predominantly in-phase, did provide support for the predictive capability of the planetary model in respect to ~ 155 day periodicity in flare activity. Prediction of Rieger periodicity with the model during solar cycles 24, 25 and 26 is therefore conditional on the north and south components varying in-phase during these solar cycles.

The possibility of forecasting differentiates the planetary model from previous hypotheses concerning Rieger periodicity, (Wolff 1992, Bai and Sturrock 1993, Lou 2000, Wang and Sheeley 2003, Zaqarashvili 2010, Gurgenashvili et al 2016, Zaqarashvili and Gurgenashvili 2018), in that the latter hypotheses do not appear to have any predictive capability with respect to the timing of solar activity.

The earliest model, Bai and Sturrock (1993), is based on the concept that the Sun may have two bands of activity rotating with a period of 25.5 days about an axis tilted with respect to the rotation axis of Sun. The bands, called "exciters", are supposed to interact with sunspot regions to stimulate flare activity at the sub-harmonic periods, 51, 76, 102, 127, 153, 178, …. n25.5 days. It is possible to find periodicity at some of these periods in individual solar cycles, e.g. (Bai 1992, Bai 1994). However, with a series of sub-harmonic periods spaced at 25.5 days, there are numerous sub-harmonics in the intermediate periodicity range between 100 and 1000 days, and the Bai and Sturrock model provides no reason why ~155 day periodicity dominates over other periods in flare activity, as indicated in Figure 6B, or why only a few periodicities e.g. 352, 396, 530 and 836 days, dominate sunspot area periodicity, as indicated in Figure 1. Thus, the Bai and Sturrock model lacks a mechanism to select the periodicities that dominate in observed solar activity from the many available subharmonics. To address this problem in respect to ~155 day periodicity Sturrock (1996) suggested the Sun contains a second element rotating at ~ 21 days on a different axis that interacts with the 25.5 day element to enhance the ~155 day sub harmonic.

Wang and Sheeley (2003) developed a purely stochastic model of sunspot emergence at time scales less than the ~ 11 year solar cycle. They showed that random emergence of magnetic flux during a solar cycle can result in activity exhibiting periodicities in the 300 to 1000 day range with the periodicities randomly varying from solar cycle to solar cycle. However, it is difficult to reconcile the observed coherence of ~155 day periodicity over several solar cycles, Figures 10A, 10B and 14 of this paper, and the observations of Bogart and Bai (1985) and Lean (1990) with a model based on random sunspot emergence. Also, a stochastic model does not explain why specific periodicities occur more frequently than others in sunspot emergence and flare data, as illustrated for example in Figures 1 and 6, nor is there any possibility with a stochastic model of predicting the occurrence of a specific periodicity unless the stochastic process is modulated by some underlying process of stable long term periodicity, as suggested by Bogart and Bai (1985).

Some models of Rieger periodicity, (Wolff 1992, Lou 2000, Zaqarashvili et al 2010, Gurgenashvili et al 2016, Zaqarashvili and Gurgenashvili 2018) propose that Rieger periodicity arises from unstable wave modes on the Sun.

Wolff (1992) related the observed periodicities in the sunspot number, during the interval 1749 – 1979, to the periods of the coupled modes of oscillation, the g-modes, of a slowly rotating star. In particular, Wolff related the observed periodicity in sunspot number at 155 days to the theoretical period, 155.4 days, (frequency 74.48 nHz), of the l = 2, l = 3 coupled g-mode of the Sun. Subsequently, Wolff and Patrone (2010) investigated mechanisms by which planetary motion might affect the Sun. It is interesting that the 155.4 day, l = 2, l = 3, coupled g-mode periodicity is very close in period to the 155.3 day periodicity that dominates intermediate range periodicity in the present planetary model, Figure 5A.

Lou (2000) derives a series of allowed hydrodynamic Rossby wave modes with a range of periods some of which are near the commonly observed periods in solar activity such as 155, 352, 530, 836, and 1979 days that correspond to the prominent periodicities indicated in Figure 2. However as the allowed mode periods are spaced by only 13 days almost any of the observed long period periodicities could be associated with an allowed mode and Lou (2000) does not suggest why specific modes are more commonly observed. Therefore, Lou's model requires some mechanism for selecting specific modes from the many, as suggested by Dimitropoulou et al. (2008).

Zaqarashvili et al (2010) derived theoretical expressions for the periods of unstable magnetic wave modes in the range 150 – 200 days. They proposed that unstable modes in sub-surface magnetic flux tubes grow in amplitude leading to enhanced magnetic buoyancy and flotation of flux tubes to the surface. The observed intermittency of periodicity was explained (Gurgenshavili et al. 2016, Zaqarashvili and Gurgenashvili 2018) as due to dependence of Rossby wave periodicity on the toroidal magnetic field strength and it was proposed that field strength changing between 20 kG and 40 kG from solar cycle to solar cycle caused periodicity in sunspot area to change between ~155 days and ~200 days. The observed dominance of ~155 day periodicity is apparently due to the prevalence of 20 kG field strength over the 40 kG field strength. Thus, in this model, predicting sunspot activity of particular periodicity depends on predicting the time variation of the solar magnetic field strength. However, Gurgenshavili et al (2016) do not suggest how magnetic field strength might be predicted. Gurgenashvili et al (2016) confine their analysis to the periodicity range between 150 and 200 days. The occurrence of activity at periods longer than 200 days or shorter than 150 days, periodicity similar to that reported, for example, by Choudhary et al (2014), Tan and Cheng (2013), and in Figures 1, 2, 6A and 6B of this paper, is dismissed as possibly having a different physical origin.

The concept of sunspots arising from unstable Rossby wave modes at the solar tacholine is not incompatible with the concept of a planetary influence as the planetary model is based on triggering the flotation of primed or unstable magnetic flux when tidal influence peaks. Also, the present planetary model develops a range of periodicities, Figures 5A and 5B, among which a few periodicities are clearly more dominant, e.g. 146, 155, 196, 352, 540, and 836 days, indicating periodicities at which triggering of unstable magnetic flux is more likely to occur. Thus the planetary model could provide the mechanism that selects specific Rossby wave modes from among the many available as being the modes more likely to be associated with periodic solar activity.

Any planetary model suffers the disadvantage of lacking a generally accepted physical mechanism by which extremely small gravitational tidal effects trigger the emergence of sunspots. Obviously, the presence of very unstable wave modes in sub-surface magnetic flux would enhance the possibility that small periodic tidal effects could trigger flux emergence. There have been proposals for amplification of tidal influence. Scafetta (2012) suggested nuclear fusion amplification of the very small tidal density variations to increase the tidal influence. Seker (2012) proposed that planetary tides may be amplified by resonance with Alfven waves on the Sun. One or both mechanisms might assist the triggering of flux emergence by tidal effects. Nevertheless, a planetary model has the significant advantage in having the potential for both short and long term predictive capability. This paper develops a simple planetary model to predict Rieger periodicity. If the predictions made using the model prove successful, this may encourage further research into planetary influence on solar activity.